# The Real Space Correlation Function Measured from the APM Galaxy Survey

C.M. Baugh.
*Department of Physics, Science Laboratories, South Road, Durham DH1 3LE*

**ABSTRACT**

We present a determination of the real space galaxy correlation function, $\xi(r)$, for galaxies in the APM Survey with $17 \leq b_J \leq 20$. We have followed two separate approaches, based upon a numerical inversion of Limber's equation. For $\Omega = 1$ and clustering that is fixed in comoving coordinates, the correlation function on scales $r \leq 4h^{-1}$Mpc is well fitted by a power law $\xi(r) = (r/4.5)^{-1.7}$. There is a shoulder in $\xi(r)$ at $4h^{-1}$Mpc $\leq r \leq 25h^{-1}$Mpc, with the correlation function rising above the quoted power law, before falling and becoming consistent with zero on scales $r \geq 40h^{-1}$Mpc. The shape of the correlation function is unchanged if we assume that clustering evolves according to linear perturbation theory; the amplitude of $\xi(r)$ increases however, with $r_0 = 5.25h^{-1}$Mpc. We compare our results against an estimate of the real space $\xi(r)$ made by Loveday *et al.* (1995a) from the Stromlo-APM Survey, obtained using a cross-correlation technique. We examine the scaling with depth of $\xi(r)$, in order to make a comparison with the shallower Stromlo-APM Survey and find that the changes in $\xi(r)$ are within the $1\sigma$ errors. The estimate of $\xi(r)$ that we obtain is smooth on large scales, allowing us to estimate the distortion in the redshift space correlation function of the Stromlo-APM Survey caused by galaxy peculiar velocities on scales where linear perturbation theory is only approximately correct. We find that $\beta = \Omega^{0.6}/b = 0.61$ with the $1\sigma$ spread $0.38 \leq \beta \leq 0.81$, for $\Omega = 1$ and clustering that is fixed in comoving coordinates; $b$ is the bias factor between fluctuations in the density and in the light. For clustering that evolves according to linear perturbation theory, we recover $\beta = 0.20$ with $1\sigma$ range $-0.02 \leq \beta \leq 0.39$. We rule out $\beta = 1$ at the $2\sigma$ level. This implies that if $\Omega = 1$, the bias parameter must have a value $b > 1$ on large scales, which disagrees with the higher order moments of counts measured in the APM Survey (Gaztañaga 1994).

## 1 INTRODUCTION

The galaxy two-point correlation function $\xi(r)$ is a powerful discriminant between models of structure formation in the universe. Analysis of recently completed galaxy surveys has shown that excess correlations are observed on scales larger than $10h^{-1}$Mpc (Maddox *et al.* 1990, Efstathiou *et al.* 1990, Saunders *et al.* 1991), compared with the predictions of the standard Cold Dark Matter model. The correlation function measured from a redshift survey of galaxy positions will be distorted by the peculiar motions of the galaxies. The correlations on large scales will be boosted by coherent flows (Kaiser 1987), whilst those on small scales are reduced by the virialised motion of galaxies within groups and clusters.

A measurement of these distortions, either by resolving the redshift space correlation function $\xi(s)$ into harmonic components (Hamilton 1992), or by comparison with the real space correlation function gives us information about the whole mass distribution, rather than just the luminous component. To take the latter approach, a reliable way of estimating the real space correlation function, particularly on large scales where linear perturbation theory applies, is needed.

The method normally used to obtain the real space correlation function is to try to invert two dimensional measures of clustering. The projected distribution of galaxies is unaffected by peculiar motions and angular catalogues usually contain orders of magnitude more galaxies than redshift catalogues.

We have described a method for recovering the real space power spectrum in three dimensions, $P(k)$, which is the Fourier transform of $\xi(r)$ (Baugh & Efstathiou 1993, 1994; hereafter BE93, BE94). We adapted an iterative technique due to Lucy (1974) to numerically invert Limber's equation (Limber 1954), which relates a measure of clustering in two dimensions, such as the angular correlation function $w(\theta)$ or the power spectrum, to an integral equation that projects a three dimensional measure of clustering. The method requires a model for the redshift distribution of the galaxies in the survey and assumptions about the evolution of clustering with redshift.

There are several advantages to be gained by using such an approach. The result that is obtained for $\xi(r)$ is independent of the initial form used to start the iterations. If a fit of $\xi(r)$ was made to reproduce the observed $w(\theta)$, the result would be restricted by the parametric form adopted; important features in correlation function could be missed unless they were anticipated beforehand in choosing a shape for $\xi(r)$. The correlation function that we recover is smooth and the method converges rapidly to a stable solution. This is because the numerical technique is robust and does not involve any potentially unstable numerical operations, such as differentiation, unlike the method of Fall & Tremaine (1977) and Parry (1977), who proposed using Mellin transforms to invert Limber's equation.

The real space correlation function can also be estimated from redshift surveys. Davis and Peebles (1983) computed the correlation function in redshift space as a function of the separation of galaxies parallel and perpendicular to the line of sight. This function is then projected and is related to $\xi(r)$ by an Abel integral which can be inverted. However, the binning of galaxy distribution into a two dimensional grid makes the correlation function noisy.

This method has been adapted for sparse-sampled redshift surveys by Saunders *et al.* (1992). In sparse-sampled surveys, galaxies are selected at random at some set rate from the parent two dimensional catalogue (*e.g.* at a rate of 1 in 6 for the QDOT survey, and 1 in 20 for the Stromlo-APM Survey). It can be shown that comparable information can be extracted about the large scale structure in the survey, as would be obtained from 1 in 1 sampling, but with less expenditure of telescope time (Kaiser 1986). By calculating the cross correlation function between the projected redshift survey and the parent catalogue, the noise is greatly reduced, allowing an inversion of the projected correlation function to be carried out.

The method requires a weighting scheme for the galaxies, which is calculated by assuming a prior form for the correlation function, though it is claimed that the weights are not that sensitive to the actual form chosen.

A power law form for the spatial correlation function can be chosen so that the projected cross-correlation function is reproduced. Alternatively the relation between the spatial and cross correlation function can be inverted directly, without having to restrict the form of the spatial correlation function. The direct inversion involes taking a derivative of the cross-correlation function, which leads to a result that is not smooth, particularly on larger scales where the estimated cross correlations are noisier.

Loveday *et al.* 1995a,b (L95a,L95b) have applied this technique to the APM-Stromlo redshift catalogue. One aim of this paper is to compare these results with those obtained from the inversion scheme of BE93. We derive the relativistic version of the integral equation relating $w(\theta)$ to $\xi(r)$, in a form that is suitable for inversion by Lucy's method in Section 2. The real-space $\xi(r)$ is computed in Section 3 by taking the Fourier transform of the results for $P(k)$ obtained by BE93. We perform the Lucy inversion in Section 4.

In order to compare our results obtained using the APM Survey for galaxies with magnitudes in the range $17 \leq b_J \leq 20$, with those of L95a for galaxies with $b_J \leq 17.15$, we need to examine the scaling of our measurement of $\xi(r)$ with survey depth, which we discuss in Section 5.

We estimate the size of peculiar velocity distortions in the Stromlo-APM Survey by comparing the redshift space correlation function of L95a with our estimate of the real space correlation function in Section 6. Finally our conclusions are presented in Section 7.

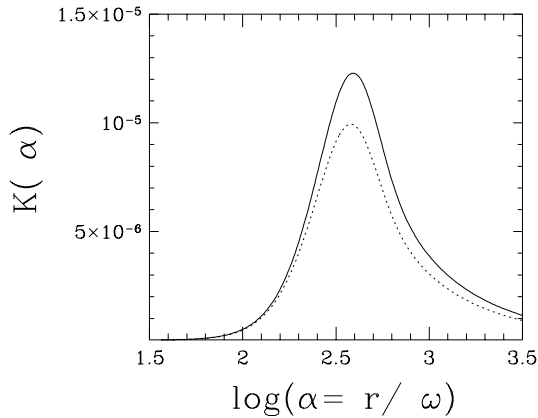

**Figure 1.** The kernel function defined by equation 9 for the case of a spatially flat universe with $\Omega = 1$ and clustering that is fixed in comoving coordinates $\alpha = 0$ (solid line). The kernel with linear evolution of clustering in comoving coordinates ($\alpha = 2$) is shown by the dotted line.

## 2 THE INTEGRAL EQUATION

The spatial correlation function $\xi(r,t)$ is related to the angular correlation function $w(\theta)$ by Limber's equation (Limber 1954), if we make the assumptions that clustering is independent of luminosity and that the correlation function tends to zero on scales approaching the depth of the survey.

We shall rewrite the relativistic version of Limber's equation (equation 1 of BE93; see also Peebles 1980, §56) in a form that is more suitable for inversion by Lucy's technique:

$$w(\varpi) = \frac{2 \int_0^\infty \int_0^\infty x^4 F^{-2} a^6 p^2(x) \xi(r,t)\, dx\, du}{\left[\int_0^\infty x^2 F^{-1} a^3 p(x) dx\right]^2}. \quad (1)$$

The co-ordinate distance $x$ is related to the physical separation between galaxy pairs that are separated by angle $\theta$ on the sky by:

$$r^2 = a^2 \left[u^2/F^2(x) + x^2 \varpi^2\right] \quad (2)$$
$$\varpi = 2\sin(\theta/2), \quad (3)$$

where the function $F(x)$ depends upon the cosmological model used (see equation 56.13 from Peebles 1980) and $u = x_1 - x_2$ (Peebles 1980 equation 56.16).

We express the selection function $p(x)$ of the survey in terms of the redshift distribution $(dN/dz)$

$$\mathcal{N} = \int \frac{x^2}{F(x)} a^3 p(x) dx = \frac{1}{\Omega_s} \int \frac{dN}{dz} dz, \quad (4)$$

where $\mathcal{N}$ is the number density of galaxies per steradian and $\Omega_s$ is the solid angle covered by the survey.

The redshift distribution is parameterised by the median redshift, $z_{b_J} = 1.412 z_c$ with

$$z_c(b_J) = \left(0.016(b_J - 17)^{1.5} + 0.046\right)/1.412, \quad (5)$$

which is a function of the apparent magnitude, $b_J \geq 17$. The redshift distribution is represented by the formula

$$dN = \frac{3\mathcal{N}\Omega_s}{2z_c^3(m)} z^2 \exp\left[-(z/z_c)^{3/2}\right] dz, \quad (6)$$

which was chosen to provide a fit to the redshift distribution of galaxies in the Stromlo-APM survey of Loveday *et al.* (1992), and the fainter surveys of Broadhurst *et al.* (1988) and Colless *et al.* (1991, 1993) (*cf* Figure 1 of BE93). Note that this formula is different to the redshift distribution that would be predicted from the Maddox *et al.* (1990) fit for the APM luminosity function (see Gaztañaga 1995). Equation 6 also provides a good fit to recently compiled Autofib data in the magnitude range $17 \leq b_J \leq 20$ (Ellis *et al.* in preparation; G. Efstathiou, private communication).

We shall approximate the evolution of the spatial correlation function measured in terms of the comoving separation between galaxy pairs as

$$\xi(r', t) = \frac{\xi(r')}{(1+z)^\alpha}. \quad (7)$$

The parameter $\alpha$ is a function of scale and epoch. However, this would make it impossible to invert the integral equation, which would then be a function of two variables, given an observed function of one variable. On small scales, if the approximation of stable clustering applies (Peebles 1980), the two-point function will grow as $1/(1+z)^3$. However, this paper is mainly concerned with determining the shape of $\xi(r)$ on large scales. In view of the low median redshift of $b_J \sim 20$ APM galaxies ($z_m \sim 0.2$ from equation 5), we shall examine the cases $\alpha = 0$, which corresponds to clustering fixed in comoving coordinates and $\alpha = 2$, which is the prediction of linear perturbation theory.

We change variables in equation 1 from $u$ to $r' = r/a$, the comoving separation. Finally, changing the order of integration in Limber's equation gives us the desired result (*cf* Peebles 1980, equation 53.1 for the non-relativistic version)

$$w(\varpi) = \frac{1}{\varpi}\int_0^\infty r\,\xi(r)\,K(r/\varpi)\,\mathrm{d}r, \qquad (8)$$

where the kernel $K(r/\varpi)$ is defined by the integral

$$K(\alpha) = \frac{2}{(\mathcal{N}\Omega_s)^2}\int_0^\alpha \frac{F(x)(\mathrm{d}N/\mathrm{d}x)^2}{[\alpha^2-x^2]^{1/2}}\frac{1}{(1+z)^\alpha}\mathrm{d}x, \qquad (9)$$

In Figure 1, we show the form of the kernel for the case of a spatially flat universe, $\Omega = 1$, with the solid line showing clustering fixed in comoving coordinates ($\alpha = 0$) and the dotted line showing clustering evolving according to linear theory ($\alpha = 2$). The mean redshift of galaxies with magnitudes in the range $b_J = 17 - 20$, predicted by our parametric form for the redshift distribution (eqn(6)) is $\bar{z} = 0.14$. This gives a reduction of $\sim 25\%$ in the amplitude of the peak of the kernel function if we assume that the clustering evolves according to linear theory.

## 3 FOURIER TRANSFORM ESTIMATE OF THE CORRELATION FUNCTION

The power spectrum of galaxy clustering is the Fourier transform of the spatial correlation function

$$\xi(r) = \frac{1}{2\pi^2}\int \mathrm{d}k\, k^2 P(k)\frac{\sin(kr)}{kr}, \qquad (10)$$

where we now use $r$ for comoving separation and $k$ is the comoving wavenumber.

We can use equation 10 to obtain an estimate of the spatial correlation function, using the three dimensional power spectrum recovered by BE93. In addition, this allows us to check the form of equations 8 and 9, by taking the spatial correlation function obtained from the Fourier transform and using it to calculate $w(\theta)$. BE93 show in their figure 8, that the power spectrum they recover predicts a $w(\theta)$ that is in excellent agreement with the APM data points (taken from Maddox *et al.* 1990).

Throughout this paper, we analyse the APM Survey split up into four roughly equal strips in right ascension (see Figure 2 of BE94). We show the Fourier transform of the three dimensional power spectrum obtained for each zone by the broken lines in Figure 2. The mean of these estimates of the spatial correlation function is shown by the triangles and the 1 sigma scatter is indicated by the error bars. We have extrapolated the form of the power spectrum as $P(k) \propto k^{-1.25}$ on small scales in the Fourier transform to obtain $\xi(r)$. Figure 2(a) shows the form of the correlation function estimated for each zone on a log-log plot up to $r = 30h^{-1}$ Mpc. The shape of the correlation function on larger scales is shown on a linear scale in Figure 2(b).

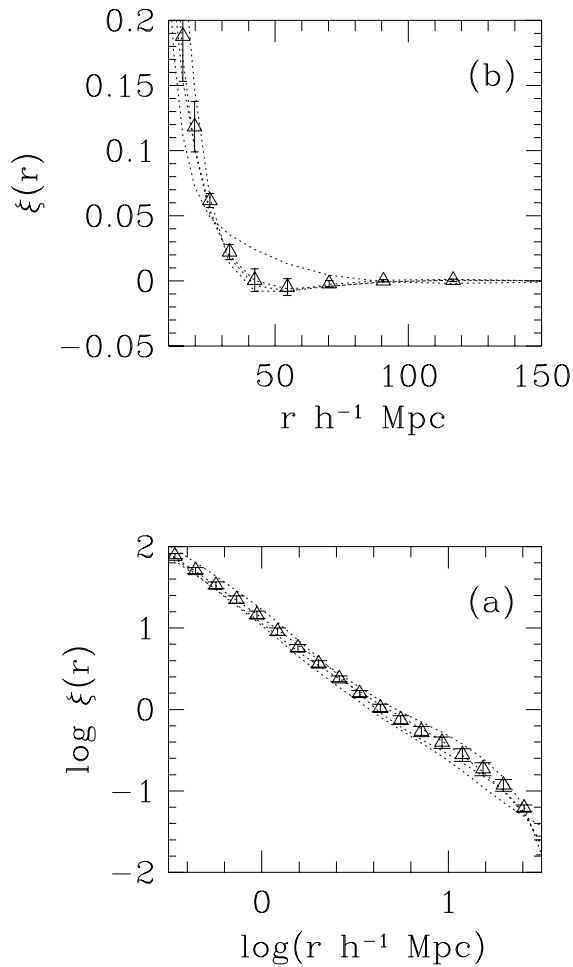

**Figure 2.** The spatial correlation function obtained by Fourier transforming the three dimensional power spectrum recovered by BE93 from the APM Survey split into four zones. The triangles show the mean of these estimates with one sigma error bars. (a) shows a log-log plot of $\xi(r)$ on scales smaller than $30h^{-1}$ Mpc. (b) shows the form of $\xi(r)$ on large scales.

Finally in this Section, we use equation 8 to compute the expected $w(\theta)$ from the correlation function obtained in each zone. The mean of the angular correlation function so obtained for each zone is shown by the solid line in Figure 3. The dashed line shows the average $w(\theta)$ computed from the power spectrum in each zone, using equation 5(a) of BE93. In this case, no extraploation of the form of the power spectrum to small scales was made, in order to show the scale at which the extrapolation is important. The points show the actual $w(\theta)$ measured for the APM Survey by Maddox et al. (1990), with $2\sigma$ error bars.

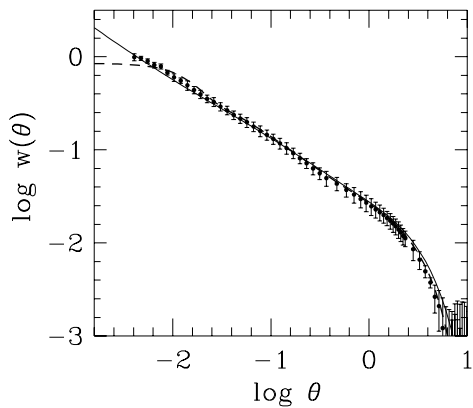

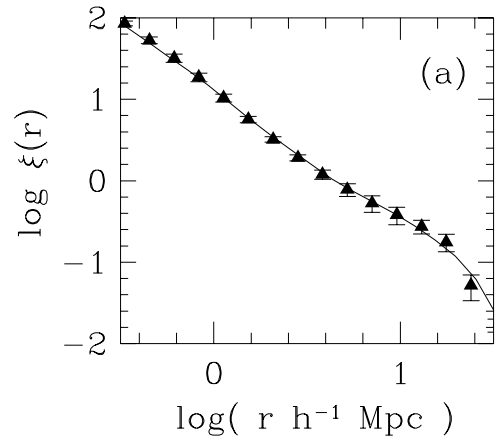

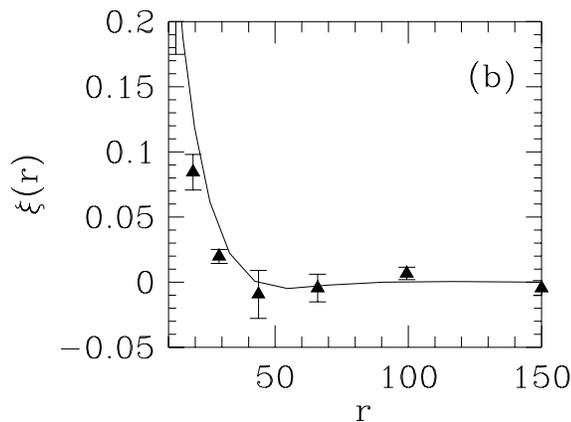

**Figure 3.** An illustration of the accuracy of the estimates of $\xi(r)$ obtained by Fourier transforming the results of BE93 for $P(k)$. The points show the APM $w(\theta)$ averaged over four zones with $2\sigma$ errors, from Maddox et al. (1990). The dashed line is the mean $w(\theta)$ obtained from $P(k)$ and the solid line is the mean angular correlation function from $\xi(r)$.

## 4 APPLICATION OF LUCY'S METHOD

Following BE93 and BE94, we can numerically invert the integral equation (equation 8) using a technique based upon Lucy's method (Lucy 1974) to recover the spatial correlation function.

Beginning with a power law form for the spatial correlation function (though the result of the inversion is not dependent upon the initial form chosen for $\xi(r)$), we compute the corresponding angular correlation function and compare it to the measured $w(\theta)$. This comparison is then used to generate a new estimate of the spatial correlation function, and the process is repeated. If the $n^{th}$ estimate of the spatial correlation function gives us an estimate of the angular correlation function as:

**Figure 4.** The spatial correlation function obtained by Fourier transforming $P(k)$, shown by the solid line, compared with the result obtained by applying Lucy's method to $w(\theta)$, shown by the points with $1\sigma$ errors. (a) shows a log-log plot of $\xi(r)$ on scales smaller than $30h^{-1}$ Mpc. (b) shows the form of $\xi(r)$ on large scales.

$$w^n(\omega_j) = \sum_i r_i/\omega_j K(r_i/\omega_j)\xi^n(r_i)\Delta r_i, \quad (11)$$

then the revised estimate of the spatial correlation function is given by

$$\xi^{n+1}(r_i) = \xi^n(r_i)\frac{\sum_j w^0(\omega_j)/w_j^n(\omega_j)r_i/\omega_j K(r_i/\omega_j)}{\sum_j r_i/\omega_j K(r_i/\omega_j)}, \quad (12)$$

where $w^0(\omega)$ is the measured angular correlation function. We have replaced the integrals by summations and typically used 60 logarithmic bins for $w(\omega)$ and 30 bins for $\xi(r)$ in the range $r = 0.001$ to $r = 150h^{-1}$ Mpc.

Again, we apply the inversion individually to each of the four zones into which we have split the APM Survey. The iterations are stopped when the $w(\omega)$ obtained from the current estimate of the spatial correlation function provides a good match to the measured angular correlation function for that particular zone.

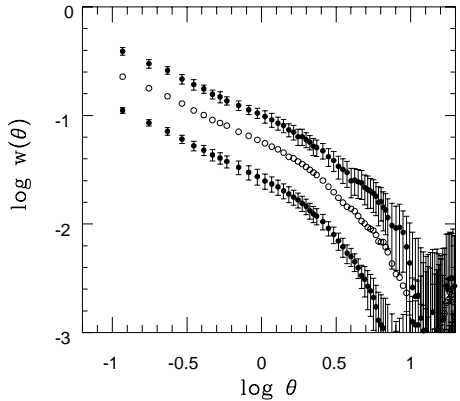

**Figure 5.** The angular correlation function measured for APM galaxies in the magnitude range (in order of decreasing amplitude); $17 \leq b_J \leq 18$, $17 \leq b_J \leq 19$ and $17 \leq b_J \leq 20$. (We have omitted the $1\sigma$ errors, obtained by averaging over zones, for the $17 \leq b_J \leq 19$ slice for clarity).

**Table 1.** Magnitude slice parameters

| $b_J$ | $\bar{z}$ | absolute mag $M_B - 5\log h$ | |
|---|---|---|---|
| 17 - 18 | 0.069 | -19.47 | -18.47 |
| 18 - 19 | 0.101 | -19.26 | -18.25 |
| 19 - 20 | 0.143 | -18.95 | -17.95 |
| 17 - 19 | 0.097 | -20.17 | -18.17 |
| 17 - 20 | 0.138 | -20.87 | -17.87 |

We plot the mean of the recovered spatial correlation functions in Figure 4 with $1\sigma$ error bars. The solid line shows the mean spatial correlation function recovered by taking the Fourier transform of the three dimensional power spectrum in each zone. The different approaches to estimating the form of the spatial correlation function are consistent to within $1\sigma$ on scales up to $\sim 20 h^{-1}$ Mpc and within $2\sigma$ up to the maximum scale for which the correlation function was recovered $150 h^{-1}$ Mpc.

## 5  SCALING WITH DEPTH OF THE CORRELATION LENGTH

The correlation length is defined as the separation at which the two point correlation function is unity $\xi(r_0) = 1$. Analysis of the CfA redshift surveys has indicated that the correlation length that characterises galaxy clustering scales with the depth of the survey. Coleman *et al.* (1988) found that the correlation length in volume limited subsamples of the CfA 1 survey increased linearly with the depth of the sample. The authors interpreted this trend as evidence for a fractal galaxy distribution. The scaling with depth of the angular correlations measured for half magnitude slices of the APM Survey argues against this conclusion (Maddox *et al.* 1990, Peebles 1993). Davis *et al.* (1988) analysed an extension of the CfA survey and found that the correlation length increased in proportion to the square root of the effective depth.

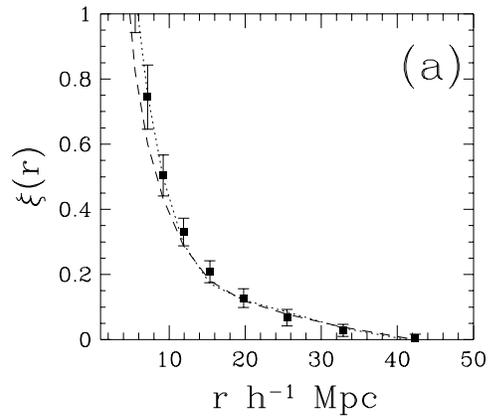

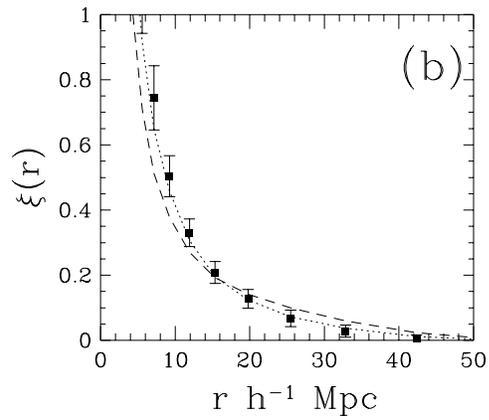

**Figure 6.** (a) The real space correlation functions obtained for APM galaxies in the magnitude range; $17 \leq b_J \leq 18$ (symbols with 1 sigma errors), $17 \leq b_J \leq 19$ (dotted line) and $17 \leq b_J \leq 20$ (dashed line). (b) $\xi(r)$ recovered from $w(\theta)$ measured from galaxies with magnitudes $17 \leq b_J \leq 18$ (symbols with 1 sigma errors), $18 \leq b_J \leq 19$ (dotted line) and $19 \leq b_J \leq 20$ (dashed line).

Two possible explanations for this be-

haviour have been put forward. The volume limited samples used in the above analyses typically have depths of $100h^{-1}$ Mpc or less, and so could be biased by density fluctuations on scale of the sample, which makes the mean density uncertain. In the CfA1 survey, the Local Supercluster biases the density to be higher than the true mean density of the universe. The APM Survey covers a much larger volume and so is not affected by local inhomogenities. Alternatively, using deeper volume limited samples includes more bright galaxies in the catalogue and there is some evidence that these galaxies are more strongly clustered than faint galaxies (Hamilton 1988, Davis *et al.* 1988, L95a,b).

In order to make a meaningful comparison of the real space $\xi(r)$ measured from the $17 \leq b_J \leq 20$ APM Survey with the redshift space correlation function obtained from the shallower Stromlo-APM Survey, we need to examine the variation of $\xi(r)$ with effective depth. Using the full $b_J = 17 - 20$ slice will give the most accurate result for $\xi(r)$ on large scales, on which the linear perturbation theory result for the ratio of redshift space to real space correlation functions will be valid (Kaiser 1987).

We have estimated the angular correlation function for the APM survey split up into several magnitude slices; figure 5 shows $w(\theta)$ for the $17 \leq b_J \leq 18$, $17 \leq b_J \leq 19$ and $17 \leq b_J \leq 20$ slices in order of decreasing amplitude. We have also calculated $w(\theta)$ for one magnitude slices: $18 \leq b_J \leq 19$ and $19 \leq b_J \leq 20$. The angular correlation function was estimated by applying the ensemble estimator to the Survey in the form of pixel maps (Maddox *et al.* 1990).

Using our parameterisation of the redshift distribution of APM galaxies, we can calculate the mean redshift of the galaxies in each slice. This gives us an indication of the typical absolute magnitude of the galaxies in each slice, shown by the range in absolute magnitude that can be seen at the mean redshift, listed in Table 1.

Maddox *et al.* (1990) demonstrate the scaling with depth of $w(\theta)$ measured from different magnitude slices. This provides a test of the 'fairness' of the survey (Peebles 1980) and of the parameterisation of the luminosity function or redshift distribution of the survey galaxies. We shall apply the analogous test here, by computing the spatial correlation function for each of the $w(\theta)$ curves described above. Given the large differences between the angular correlations before any scaling is applied in Figure 5, the recovery of a universal form for $\xi(r)$ is a stong test of our parameterisation of the redshift distribution and our assumptions about the evolution of clustering.

We recover $P(k)$ for each slice using the method of BE93, and then take the Fourier transform to obtain $\xi(r)$. The results are shown in Figures 6 where we have assumed that the clustering is fixed in comoving co-ordinates.

The results for the brightest slice are shown by the points with the $1\sigma$ errors, with $\xi(r)$ for the other slices shown by the lines. The $\xi(r)$ recovered from each magnitude slice of the APM Survey are consistent within one sigma errors, especially for the broader magnitude slices.

There are several explanations for the small discrepancy between the correlation function recovered from the various magnitude slices. The $w(\theta)$ used are uncorrected for the merging of images. For the $17 \leq b_J \leq 20$ slice, this correction is estimated to be in the region of a boost in amplitude of $\sim 15\%$ (Maddox *et al.* 1990, Gaztañaga 1994). The corrections for each magnitude slice will be different. However, the best correction to make would be to each zone individually, rather than adjusting the mean of the $w(\theta)$, as the two zones nearest the Galactic plane are affected the most. For the narrow magnitude slices in Figure 6(b), the parametric form that that we have adopted for the redshift distribution may not give the best representation of distribution for the slice. Finally, we could be seeing the effects of luminosity segregation on the amplitude of the correlation function. L95a, b have shown that there is some evidence for the faintest galaxies in the Stromlo-APM survey being less strongly clustered than galaxies of the characteristic luminosity $M_* = -19.50$ and brighter. This affect is cleaner in volume limited samples however, where the distance cut corresponds to a cut in absolute magnitude.

## 6 ESTIMATING REDSHIFT SPACE DISTORTIONS

Peculiar motions of galaxies distort the pattern of clustering when the galaxy positions are mapped using distances derived from their redshifts. The direction averaged correlation function in redshift space, $\xi(s)$, is related to the real space correlation function, $\xi(r)$, on scales where linear perturbation theory is valid by (Kaiser 1987)

$$\frac{\xi(s)}{\xi(r)} = 1 + 2/3 \ \beta + 1/5 \ \beta^2, \qquad (13)$$

where $\beta = \Omega^{0.6}/b$, and $b$ is the 'bias' parameter relating the fluctuations in mass to those in the galaxy distribution $\left(\frac{\delta\rho}{\rho}\right)_g = b \left(\frac{\delta\rho}{\rho}\right)_m$.

Ideally, one would like to apply this formula on the largest scales possible, within the limit of the depth of the survey, in order to estimate the size of the peculiar velocity distortions. However, the estimates of the redshift space correlation function from the Stromlo-APM Survey become noisy on scales $r > 30 h^{-1}$ Mpc.

Table 2. Redshift space distortions measured from APM Surveys

| Source/Model | $\beta = \Omega^{0.6}/b$ | $1\sigma$ range | $2\sigma$ range |
|---|---|---|---|
| Tadros et al. (95a) | 0.55 | 0.21 - 0.89 | - |
| L95b | 0.36 | - | -0.03 - 0.75 |
| $\Omega = 1, \alpha = 0$ | 0.61 | 0.38 - 0.81 | 0.15 - 0.99 |
| $\Omega = 1, \alpha = 2$ | 0.20 | -0.02 - 0.39 | -0.26 - 0.57 |
| $\Omega = 0.2, \alpha = 0$ | 0.45 | 0.22 - 0.66 | -0.02 - 0.83 |

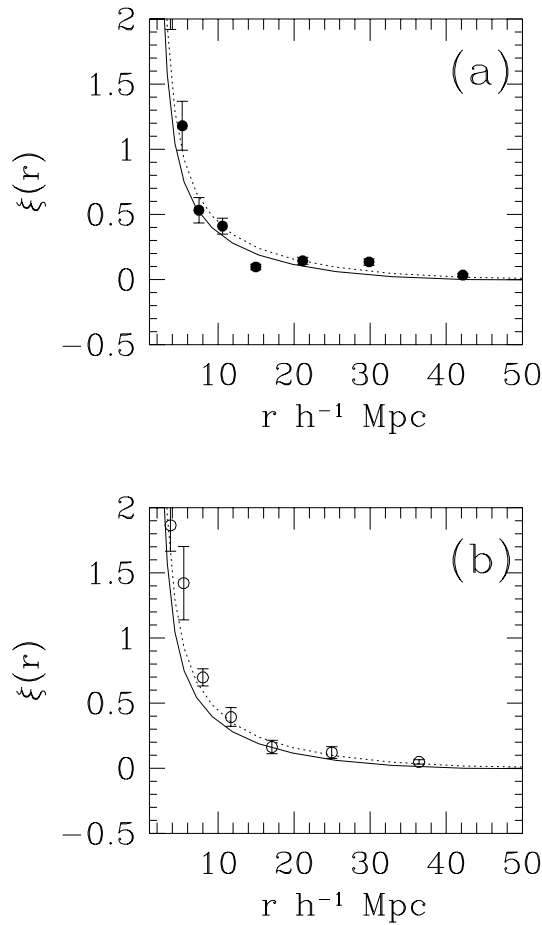

**Figure 7.** The APM Survey $\xi(r)$ compared with the correlation function measured by L95a,b for the APM Stromlo Survey. The lines correspond to the real space correlation function: the solid line is for $\Omega = 1$, $\alpha = 0$; the dotted line is for $\Omega = 1$, $\alpha = 2$. The points in (a) are the real space $\xi(r)$ for the Stromlo-APM Survey, the points in (b) are the redshift space correlation function. The error bars are from nine bootstrap resamplings of the redshift survey.

The range of scales on which linear perturbation theory is valid has been studied using the results of N-body simulations by Baugh & Efstathiou (1994) and Baugh, Gaztañaga & Efstathiou 1995. These authors found that mildly nonlinear evolution of the density field causes a transfer of power from large to small scales, for CDM like power spectra. On scales between $10 - 30 h^{-1}$ Mpc the size of this effect was typically to reduce the correlations by a factor of $\sim \leq 10\%$.

A further nonlinear effect is the small scale damping of correlations in redshift space caused by virialised motions. Over some range of scales, which can best be determined from N-body simulations that have a similar power spectrum to that observed (Tadros & Efstathiou 1995b), there will be a mixture of this damping effect and the boost in amplitude predicted by equation 13. In this paper, we will examine the ratio of the redshift space to real space correlation functions over the range $8 \leq r \leq 25 h^{-1}$ Mpc, and make the approximation that the damping due to virialised motions is negligible over these scales. This is larger than the scales used by L95b (who used $5 - 12 h^{-1}$ Mpc), because our estimate of the real space correlation function is smoother and less noisy on larger scales. However, Tadros & Efstathiou 1995a,b have made an estimate of $\beta$ by measuring the redshift space power spectrum for the Stromlo-APM Survey, in comparison with the $P(k)$ of BE93, on much larger scales, $\lambda = 2\pi/k = 53 - 120 h^{-1}$ Mpc.

We compare the Stromlo-APM Survey correlation functions, with the real space correlation function measured from the APM Survey in Figure 7, on the scales where we examine the redshift space distortions. The solid lines in the figure show the real space $\xi(r)$ estimated for $\Omega = 1$ and with clustering fixed in comoving coordinates: the dashed lines show $\xi(r)$ for clustering that evolves according to linear perturbation theory ($\alpha = 2$).

Figure 7(a) shows the real space $\xi(r)$ estimated by L95a, using the cross-correlation technique of Saunders *et al.* (1992), plotted as filled points with bootstrap errors. Figure 7(b) shows the redshift space correlation function.

To determine the ratio between the redshift space and real space correlation function, we have taken the Fourier transfrom of the APM Survey $P(k)$ at the values of $r$ for which L95a measured $\xi(s)$. Adding in quadrature the bootstrap errors on $\xi(s)$ and the zone dispersion errors on $\xi(r)$, we estimate the value of $\beta$ that minimises $\chi^2$ and give errors corresponding to $\Delta\chi^2 = 1$ ($1\sigma$) and $\Delta\chi^2 = 4$ ($2\sigma$).

We compare our results with those of L95b and Tadros & Efstathiou 1995a in Table 2. We give results for different assumptions about the evolution of clustering. Note that L95a,b have assumed that $\Omega = 1$ in calculating distances and absolute magnitudes for the Stromlo-APM galaxies. The values of $\beta$ given in table 2 are consistent within $1\sigma$. However, there are additional uncertainties in our estimate of $\beta$. We have not corrected $w(\theta)$ for contamination by merged images, which could increase the amplitude of the real space $\xi(r)$ by 15%, and thus would lead to lower estimates of $\beta$. Also, Gaztañaga (1995) has shown that different forms for the APM Survey luminosity function can lead to a 10% uncertainty in the value of $\sigma_8$.

## 7 SUMMARY

We have estimated the real space correlation function by taking the Fourier transform of the APM power spectrum, measured by BE93 and also by applying Lucy's method directly to the relativistic form of Limber's equation, relating $w(\theta)$ to $\xi(r)$.

We have compared these results to the real space $\xi(r)$ obtained from the shallower Stromlo-APM Survey using a technique developed by Saunders *et al.* (1992). The form of the correlation function that we recover on large scales is smoother than that resulting from the cross-correlation analysis. For $\Omega = 1$ and clustering that is fixed in comoving coordinates, the correlation function on scales $r \leq 4h^{-1}\mathrm{Mpc}$ is well fitted by a power law $\xi(r) = (r/4.1)^{-1.7}$. There is a shoulder in $\xi(r)$ at $4h^{-1}\mathrm{Mpc} \leq r \leq 25h^{-1}\mathrm{Mpc}$, with the correlation function rising above the quoted power law. The real space $\xi(r)$ becomes consistent with zero, within the $1\sigma$ errors on scales greater than $r \sim 40h^{-1}\mathrm{Mpc}$. If we make the assumption that clustering evolves according to linear perturbation theory, the shape of $\xi(r)$ is unchanged, though the amplitude is increased, with the the correlation length $r_0 = 5.25h^{-1}\mathrm{Mpc}$.

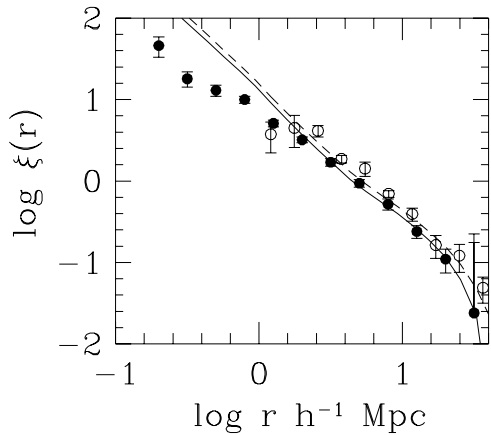

**Figure 8.** The real space $\xi(r)$ for $\Omega = 1$, with $\alpha = 0$ (solid line) and $\alpha = 2$ (dashed line) compared with the redshift space correlation function of the $1.2Jy$ survey (filled circles) and the Stromlo-APM survey (open circles).

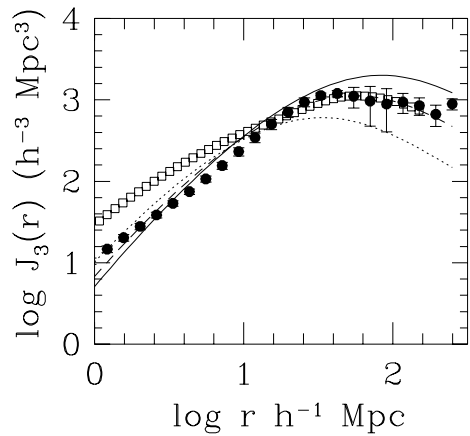

**Figure 9.** The second moment of the correlation function computed using the $\xi(r)$ recovered for each zone of the APM Survey. The filled circles show the mean of these estimates, with $1\sigma$ error bars. The cuvres show the prediction of CDM models, all normalised to give $\sigma_8 = 1$: $\Gamma = \Omega h = 0.2$ (solid line), $\Gamma = 0.3$ (dashed line); $\Gamma = 0.5$ (dotted line). The open squares show a calculation of $J_3$ for the $\Gamma = 0.3$ case using the nonlinear correlation function.

The value of $\beta$ that is measured in this paper agrees within the $1\sigma$ errors with the other published values for APM/Stromlo-APM Survey analyses. With the lowest amplitude estimate that we obtain for $\xi(r)$, namely assuming that clustering is fixed in comoving coordinates and without applying any corrections to $w(\theta)$ for merging of images, we still rule out $\beta = \Omega^{0.6}/b = 1$ at the $2\sigma$ level. Hence, to retain an $\Omega = 1$ universe, a value of the bias parameter $b > 1$ is required on large scales, which is not supported by measurements of the higher order moments of counts in cells of APM galaxies (Gaztañaga 1994).

A detailed presentation of how the three dimensional $P(k)$ compares with theoretical models and with the analysis of redshift surveys was presented in BE93. Here we restrict ourselves to a comparison of our results with the $1.2Jy$ IRAS Survey (Fisher et al. 1994) and the Stromlo-APM Survey in Figure 8. The solid line gives the real space correlation function that we obtain for $\Omega = 1$ with clustering fixed in comoving coordinates; the dashed line shows the result for clustering that evolves according to linear perturbation theory. The filled circles give the redshift space correlation function for the fully sampled $1.2Jy$ IRAS survey of 5313 galaxies from Table 2 of Fisher et al. 1994. The open circles show $\xi(s)$ for Stromlo-APM redshift survey (L95a,b) This shows the bias in clustering between optical and infrared selected galaxies. The Stromlo Survey is sparsely sampled and so does not extend to as small scales as the fully sampled $1.2Jy$ survey. The excess correlations on small scales in real space compared with redshift space are clearly apparent. On larger scales, the enhancement of clustering in redshift space is less apparent, particularly in view of the uncertainties in the amplitude of the real space correlation function. These correction would generally tend to increase the amplitude of $\xi(r)$, reducing the value of $\beta$.

We also give a calculation of the second moment of the correlation function $J_3(r) = \int_0^r r^2 \xi(r) dr$, which is used to normalise theoretical models for structure formation and is used to give the minimum variance weighting when calculating the correlation function in flux limited redshift surveys. We have calculated $J_3(r)$ using $\xi(r)$ measured for each zone of the APM Survey. The points in Figure 9 show the mean of these estimates with $1\sigma$ error bars. The curves show the predictions of CDM like models, specified by the shape parameter $\Gamma = \Omega h$ (see e.g. Efstathiou 1995). The theoretical curves have all been normalised so that the variance measured in spheres of radius $8h^{-1}$Mpc is unity. The solid line shows the form of $J_3(r)$ for a linear perturbation theory CDM power spectrum characterised by $\Gamma = 0.2$; the dashed line is for $\Gamma = 0.3$ and the dotted line gives the standard CDM model $\Gamma = 0.5$. The estimates of $J_3(r)$ on large scales reflect the uncertainty in $\xi(r)$ on these scales, enhanced by the $r^2$ weighting. The open squares show the result of a calculation of $J_3(r)$ using an approximation to the non linear shape of the two point correlation function. This was obtained by applying the linear to nonlinear transformation for the power spectrum described by Jain, Mo and White (1995).

One can see that $J_3(r)$ is insensitve to non-linear evolution on large scales. On small scales, there is disagreement between the shape of $J_3$ for the $\Gamma = 0.3$ model and the observations. There are a number of possible explanations for this:(i) there is evidence that a larger value of the clustering evolution parameter is more appropriate on small scales; (ii) some form of linear bias should be applied to the galaxy correlation function: as mentioned above the higher order moments of counts in the APM Survey do not support this on large scales (Gaztañaga 1994); (iii) CDM like models specified by the $\Gamma$ parameter have the wrong shape; (iv) some form of scale dependent bias factor is necessary.

## ACKNOWLEDGEMENTS

We would like to thank Jon Loveday for providing the APM Stromlo Survey correlation functions in electronic form, and George Efstathiou, Enrique Gaztañaga and Helen Tadros for discussions and comments on the manuscript. We also thank Andrew Hamilton for a detailed and helpful referee's report. The author is supported by a PPARC Research Assistantship.